\documentclass[12pt,preprint]{aastex}
\usepackage{graphicx,natbib}
\usepackage{emulateapj5}

\begin{document} 

\title{{Halos of Spiral Galaxies. I. The Tip of the Red Giant Branch as a 
        Distance Indicator\altaffilmark{1}}}

\author{M. Mouhcine\altaffilmark{2,3}, 
        H.C. Ferguson\altaffilmark{4},
        R.M. Rich\altaffilmark{2}, 
        T.M. Brown\altaffilmark{4}, 
        T.E. Smith\altaffilmark{4}}
\altaffiltext{1}{Based on observations with the NASA/ESA Hubble Space 
                 Telescope, obtained at the Space Telescope Science Institute,
                 which is operated by the Association of Universities
                 for Research in Astronomy, Inc.,under NASA contract 
                 NAS 5-26555}
\altaffiltext{2}{Department of Physics and Astronomy, UCLA, Math-Science 
                  Building, 8979, Los Angeles, CA 90095-1562}
\altaffiltext{3}{Present address: School of Physics and Astronomy, 
                 University of Nottingham, 
                 University Park,  Nottingham NG7 2RD, UK}
\altaffiltext{4}{Space Telescope Science Institute, 3700, San Martin Drive,
                 Baltimore, MD, 21218, USA}

\begin{abstract}
We have imaged the halo populations of a sample of nearby spiral galaxies 
using the Wide Field Planetary Camera 2 on broad the Hubble Space Telescope 
with the aim of studying the stellar population properties and relating 
them to those of the host galaxies. In four galaxies, the red-giant 
branch is sufficiently well populated to measure the magnitude of the 
tip of the red-giant branch (TRGB), a well-known distance indicator.  
Using both the Sobel edge-detection technique and maximum-likelihood 
analysis to measure the $I$-band magnitude of the red giant branch tip, 
we determine distances to four nearby galaxies: NGC~253, NGC~4244, and 
NGC~4945, NGC~4258. For the first three galaxies, the TRGB distance 
determined here is more direct, and likely to be more accurate, than 
previous distance estimates. In the case of NGC~4258, our TRGB distance 
is in good agreement with the the geometrical maser distance, supporting 
the the Large Magellanic Cloud distance modulus $(m-M)_0 = 18.50$ that 
is generally adopted in recent estimates of the Hubble constant.

\end{abstract}


\section{ Introduction}

\subsection{ Spiral galaxy Halos}

The diffuse stellar halo component of galaxies represents a tiny fraction 
of the mass (about 1\% in the case of the Milky Way; Morrison 1993) and 
the low surface brightness of the population makes study in extragalactic 
settings difficult. However, galactic halos are unique laboratories for 
investigating fundamental galaxy properties. Age and/or metallicity
distributions of halo stellar populations and their kinematics provide
fossilized glimpses of the earliest conditions of galaxy formation. 
Halo populations may give us answers to key questions about the chronology 
of galaxy formation: how the halo formation was related to the assembly 
of galactic mass and the formation of galaxies.

Different views have been advanced toward understanding the formation 
of galactic stellar halos. The first scenario was proposed by Eggen, 
Lynden-Bell, \& Sandage (1962), where the authors proposed that the 
metal-poor stars in the Galactic halo were formed during a rapid 
collapse of a relatively uniform, isolated protogalactic cloud. 
This picture is generally viewed as advocating a rapid dissipative 
and monolithic collapse of protogalaxies. 

An increasing number of observational findings, both at high redshift 
and in nearby galaxies, together with cosmological models of structure
formation, challenge this general view, and suggest an alternative 
picture of how halos of galaxies may have formed. This alternative
paradigm for halo formation is based on the idea that the galactic 
halo was formed via the accretion, over an extended period, of small 
metal-poor satellites which underwent independent chemical evolution 
before being accreted (Searle \& Zinn 1978; see also Freeman 1987). 
The disruption of globular clusters, revealed by tidal tails 
(Grillmair et al. 1995), may also contribute to the stellar halo 
(Aguilar et al. 1988). This is supported by the observation that 
halo field stars and globular clusters in the Milky Way have similar 
mean metallicities (Carney 1993). Note that this is not the case 
for M~31 stellar halo (Durrell et al. 2001, Perrett et al. 2002). 
An interesting consequence of this hypothesis is that a fraction of
globular clusters may be stripped relics of highly nucleated dwarf
satellites (see Freeman \& Bland-Hawthorn 2002 for a detailed 
discussion).
The discovery of the Sgr dwarf galaxy (Ibata et al. 1994), streams 
and moving groups in halos of both the Galaxy (Yanny et al. 2000; 
Ivezic et al. 2000; Dohm-Palmer et al. 2001) and M31 (Ibata et al. 
2001) encourage renewed interest in halo formation through accretion 
of smaller protogalactic structures. 

Although great effort  has been dedicated in recent years to 
investigate stellar populations and/or kinematic of Local Group 
spiral halos, it is unlikely that the problem of galactic halo 
formation may be understood entirely by studies of these galaxies 
alone. Questions about the universality of halo formation mechanism(s), 
the similarity of halo global properties and their sensitivity to 
disk and/or bulge properties, among others, will require a larger 
sample of galaxies.

One means of addressing these questions is to extend the study of 
galaxy halos to other spiral galaxies. Such studies are now within 
reach of the HST. Our purpose is to study systematically the {\it
global properties of population II  halos} of nearby normal spiral 
galaxies. Such a program may reveal whether there was a universal 
formation mechanism, or whether there are mixture of different 
stellar halos; if the halos are dominated by formation of the 
nuclear bulge and perhaps enriched by gas outflow, or composed 
of debris of disintegrated dwarf galaxies revealed by the presence 
of a large fraction of intermediate age stellar populations. 
Also of great interest is the investigation of whether halo and 
disk/bulge properties are correlated.

In order to interpret the color-magnitude diagrams (CMDs) of resolved 
stars in nearby galaxies it is important to have reliable distance 
estimates. Fortunately, sometimes the CMDs themselves provide such an 
estimate.

\subsection{A distance indicator to nearby galaxies: the tip of the red
                giant branch}

Stars on the first-ascent red giant branch (RGB) climb this phase with an 
expanding convective envelope and an hydrogen burning shell. While evolving 
through the RGB, low-mass stars develop an electron-degenerate core, which 
causes an explosive start of the core-He burning phase, the so-called 
He-flash, almost independently from the initial stellar parameters, such 
as initial mass or abundance (Chiosi et al. 1992). The He-flash is followed 
by a sudden decrease of the stellar luminosity due to the extinction of 
the H-burning shell. Hence low-mass stars will accumulate along the RGB, 
reaching their maximum luminosity, during this phase, at the tip of RGB 
(TRGB). This behavior translates to an abrupt discontinuity in the 
luminosity function (Renzini 1992). As the evolution of the RGB stars 
depends essentially on the core-mass, the number of stars per luminosity 
interval will be related directly to the He-core growth rate (Iben 
\& Renzini 1983). Theoretical and observational investigations find the 
luminosity function of RGB stars to follow a simple power-law (see Zoccali 
\& Piotto 2000 and references therein). The TRGB magnitude is the location 
where the RGB luminosity function truncates. Below metallicities ${\rm 
[Fe/H] \approx -0.7}$, the location of the TRGB in the $I$ band is 
expected to be quite insensitive to age and/or metallicity (e.g., Salaris 
\& Cassisi 1997). At higher metallicities, bolometric corrections in the 
$I$ band become important and the TRGB magnitude becomes increasingly
sensitive to metallicity (e.g., Salaris \& Girardi 2005). 
 
The TRGB method is also well supported observationally. The location 
of the absolute I-band TRGB luminosity for a sample of Galactic 
globular clusters spanning a large range of abundances, ${\rm 
-2.1 \leq [\rm Fe/H] \leq -0.7}$, over ages spanning 2-15\,Gyr 
(Da Costa \& Armandroff 1990), is quite stable and insensitive to 
age and metallicity (for $[\rm Fe/H] \leq -0.7$), changing by less 
than $\sim\,0.1\,$mag (Lee et al. 1993).

Good agreement is generally found between the distances obtained using 
the TRGB method and classical distance indicators such as Cepheid and 
RR Lyrae variables (e.g., Sakai et al. 1996, Ferrarese et al. 2000). 
Being a population II distance indicator, unlike the Cepheid method 
(population I distance indicator), the TRGB method can be applied to 
any class of galaxies, making this an attractive and easy technique 
to use to estimate galaxy distances. 

Because the outer regions of galaxies are devoid of significant 
concentrations of gas and dust and the stellar density is low, studies 
of standard candles in the halo provide a means of estimating distances 
free of assumptions concerning internal reddening in galaxies or 
crowding. In \S~\ref{Data}, the details of observations using the HST 
Wide Field Planetary Camera 2 (WFPC2) and data reduction are reported. 
In \S~\ref{methods}, we present the detection of RGB stars and the 
methods for measuring the TRGB using the $I$-band luminosity function 
that we have employed. Finally, in \S~\ref{dist}, we present the 
color--magnitude diagrams, luminosity functions, and derived distances 
for the galaxies studied. 

\section{Data}
\label{Data}

\begin{figure}
\includegraphics[height=2.5in]{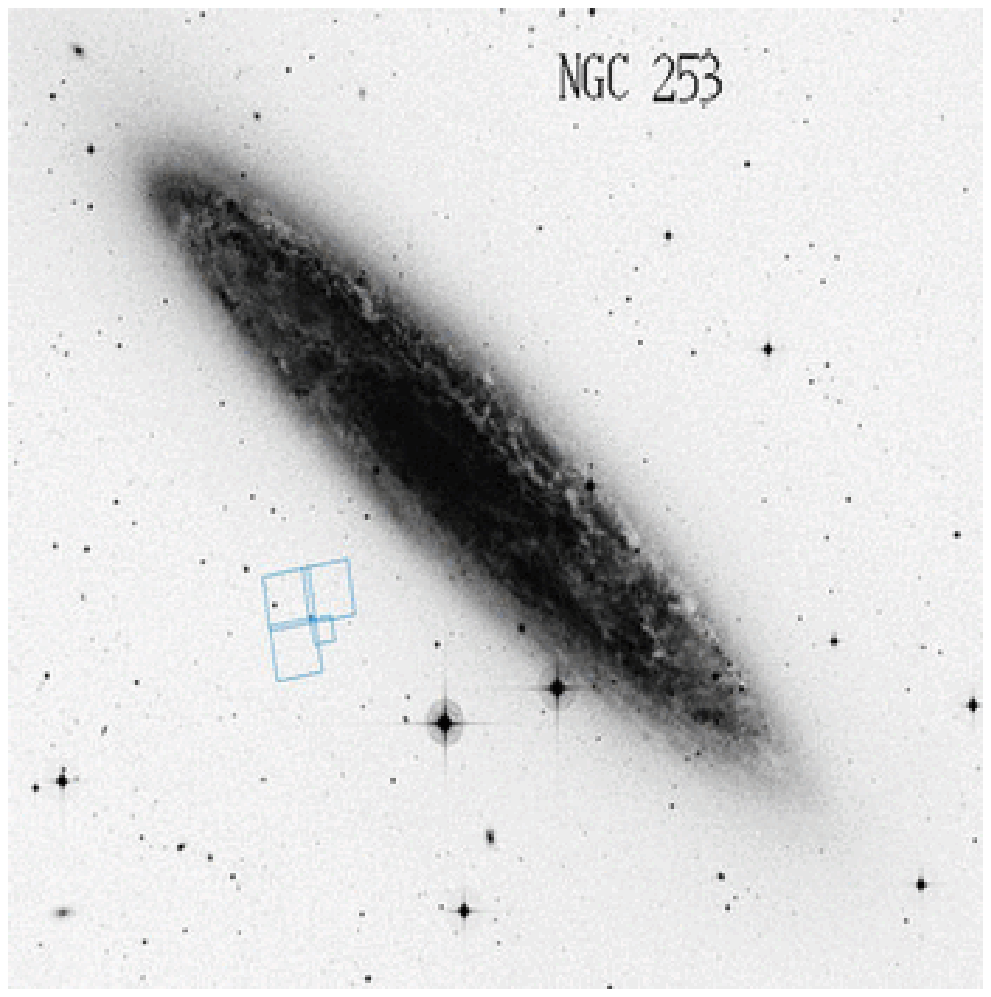}
\includegraphics[height=2.5in]{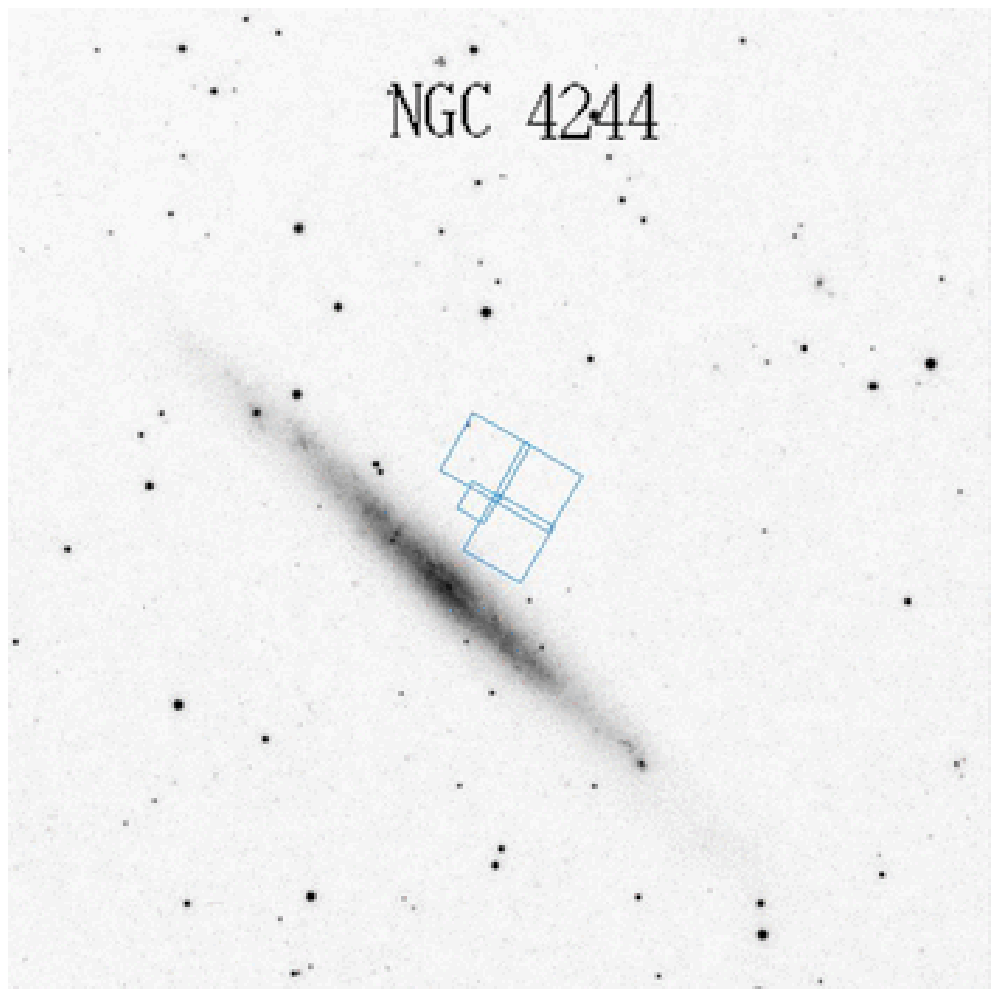}
\includegraphics[height=2.5in]{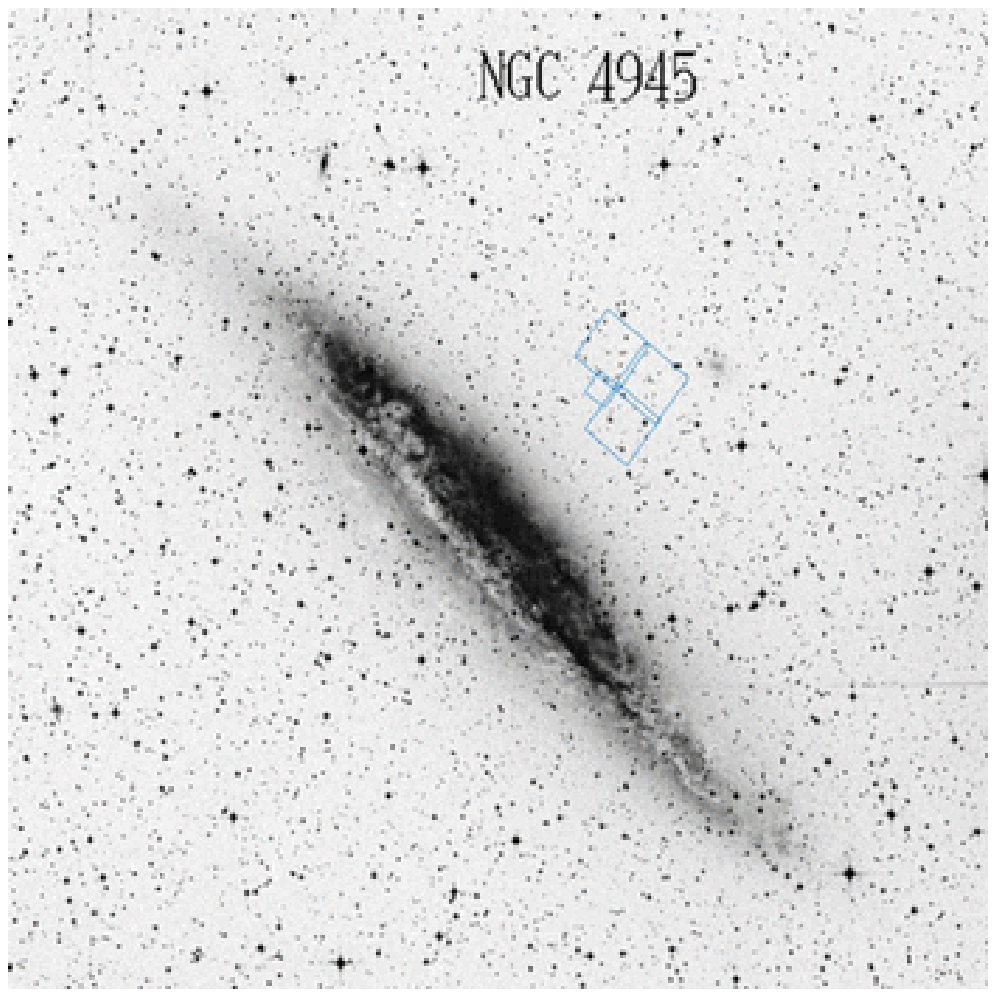}
\includegraphics[height=2.5in]{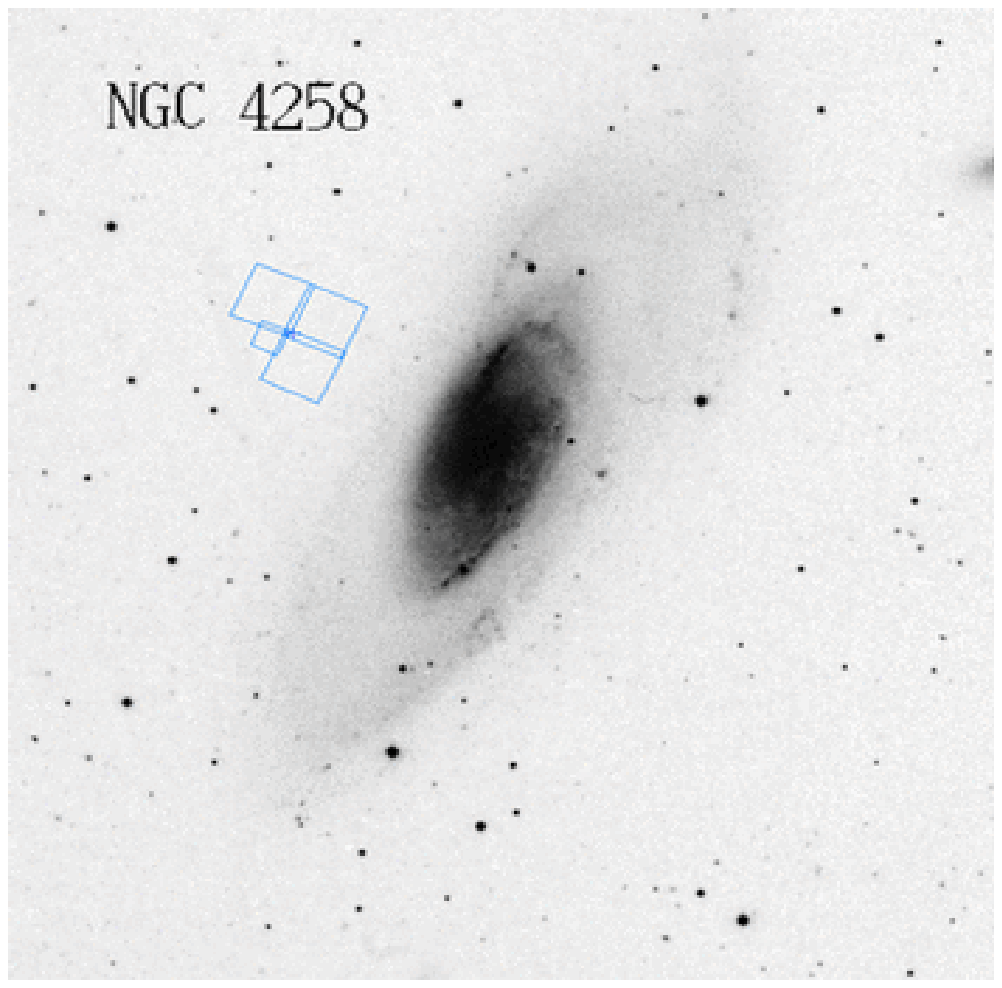}
\caption{Locations of the observed halo fields being studied  
in the paper.}
\label{field_location}
\end{figure}

Until recently, no spiral galaxy halo outside of the Local Group 
had been imaged to a depth sufficient to permit study of the halo 
metallicity distribution function or other stellar population studies, 
such as the presence of intermediate-age stars. HST observations have 
begun to change this. We have observed a sample of spiral galaxies 
with the aim of unambiguously resolving their stellar halos down 
to one or two magnitudes below the TRGB. Our selection criteria of 
the sample galaxies are as follows. We seek spiral galaxies with 
(i) morphological types between Sa to Sc, (ii) distance moduli 
$(m-M)_o<29.5$, (iii) high inclinations, i.e., $i>45$ degrees, to 
reduce any contamination from the outer disk stars, and (iv) absolute
magnitudes $M_B<-18$. The resulting sample of eight galaxies
represents most of the nearby luminous inclined spiral galaxies. 
In this paper, we estimate the distances to four galaxies in our 
sample, i.e., NGC~253, NGC~4244, NGC~4258, and NGC~4945, using 
the tip of the red giant branch technique. Fig. \ref{field_location} 
shows the locations of the observed halo fields, superimposed on 
the Digitized Sky Survey images of these galaxies. 
Madore \& Freedman (1995) argue that in order to minimize potential 
biases and to securely identify the TRGB, there should be at least 
50--100 stars in the first magnitude interval of the RGB star 
luminosity function. 
With sparser data sets the distance modulus tends to be overestimated.
Four galaxies in our sample have luminosity functions that are too
poorly populated to yield reliable TRGB distances: NGC~55, NGC~247, 
NGC~300, and NGC~3031. Hence these galaxies are not discussed in the 
present paper; the color-magnitude diagrams of the observed halo fields 
for these galaxies are shown and discussed in paper III of this series. 
Tikhonov, Galazutdinova \& Drozdovsky (2005) estimate TRGB distances
for three of these galaxies using larger HST data sets.
%

The galaxy sample is summarized in Table 1. The galaxies were observed 
through the F814W and F606W filters.  Exposure times through the F814W 
filter were set to reach S/N = 5 in the WF camera for an absolute 
magnitude ${\rm M_I = -1}$ (for galaxies with $(m-M)_0 < 27$) or 
${\rm M_I = -2}$ (for NGC4258 and NGC4945, which are more distant). 
The F606W exposure times were set to reach the same S/N at the same 
absolute magnitude for metal-poor RGB stars. The exposures were 
typically dithered over 3 separate pointings (the range being from 
2 to 6) with a pattern extending over approximately 0.25 arcseconds 
to allow rejection of hot pixels and detector artifacts.

The images were reduced through the standard HST pipeline, using the
latest flatfield observations and using contemporaneous super-dark
reference frames. The dithered frames were combined with iterative
cosmic ray rejection using software in the stsdas.dither package based
on the drizzle algorithm of Fruchter and Hook (2002). Briefly, the
images were drizzled (shifted, geometrically corrected, and resampled)
to a common frame to construct a median image reasonably free of cosmic
rays. This image was geometrically transformed back to the pixel grids
of the original image and used as the truth image for identifying 
cosmic rays. The images were then re-drizzled onto a pixel scale of
0.1$^{\prime\prime}$, masking out the cosmic rays, hot pixels, and
detector artifacts, using optimal weights based on the counts at the
sky level in each image.  A value pixfrac=0.8 was used to minimize 
the degradation of the image quality due to the resampling.

The stellar magnitudes were measured through circular apertures 
with a radius of 0.15\arcsec. These aperture magnitudes were corrected
to total magnitudes using TinyTim model (Krist 2004) point-spread 
functions. Aperture-corrections and zeropoints are applied separately
for each CCD chip. Magnitudes are in the Vega system (i.e., Vega has
magnitude=0 through both filters, when measured through a 3\arcsec
aperture). 
Diffuse sources have been excluded from the catalogs based on 
the magnitude difference between measurements through 0.09\arcsec
and 0.3\arcsec apertures. Objects where this difference is greater
than 1.3 magnitudes are generally galaxies or blends of stars. 
This demarcation was determined by starting with a Tiny-Tim 
prediction of the ratio of fluxes for a point source measured in 
those apertures. We mark objects in the images whose flux growth
exceeds that prediction and iterate to arrive at a value which 
preserves point sources while excluding those with extended radial 
profiles.

The stellar photometry was corrected for charge-transfer efficiency
(CTE hereafter) effects using the 31 May 2002 version of the Dolphin 
(2000) equations.
The CTE correction amounts to a maximum of $\sim 0.2$ mag for faint 
stars at the top of the chip in fields with the highest background, 
such as NGC3031, while reaching $\sim 0.5$ mag in fields with low sky 
background, such as NGC55. Typically the maximum correction for the 
data is $-0.5$ mag.  

\begin{figure}
\includegraphics[height=3.5in,width=3.5in]{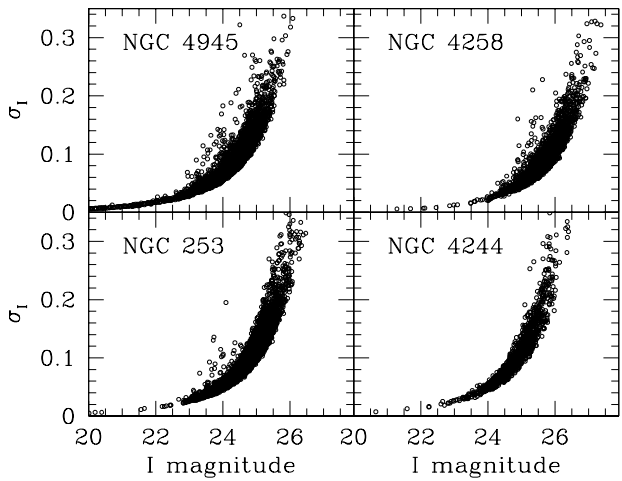}
\caption{Evolution of photometric errors (in magnitude) 
as a function of the apparent I-band magnitude.}
\label{phot_error}
\end{figure}

Transformation of the instrumental magnitudes to standard V and I 
magnitudes followed the prescriptions of Holtzman et al. (1995). 
First, we have corrected from the foreground extinction using 
${\rm A_{606W} = 2.677 \times E(B-V)}$ and 
${\rm A_{814W} = 1.815 \times E(B-V)}$, then given the instrumental 
magnitudes, we have determined the standard Johnson-Cousins 
magnitudes by solving iteratively the second degree equations 
relating the WFPC2-to-VI magnitudes using the coefficients tabulated 
by Holtzman et al. (1995). 
Fig.\,\ref{phot_error} shows the evolution of the resulting photometric 
errors for I-band as a function of the apparent magnitude. The typical 
trend of increasing photometric errors with apparent magnitude can be
seen, where most of the stars observed in the fields are affected by 
photometric errors of a few tenths of a magnitude.

\section{ Measuring the TRGB}
\label{methods}

To avoid stellar populations that may bias the analysis of the 
luminosity function, we have restricted our analysis to stars 
that match the location of the RGB sequence of metal-poor stellar 
populations. To do so, we have retained only stars that are bluer 
than a certain cut. 
For all the galaxies discussed in this paper, the remaining number 
of red giant stars after the color cut is sufficient for a reliable 
TRGB estimate by the criterion of Madore \& Freedman (1995); i.e., 
there are more than 100 stars in the one-magnitude interval fainter 
than the TRGB.
It is worth mentioning that as long as the aim is measurement of 
the TRGB, there is no need to include {\it all} the observed stars 
to construct the luminosity function. When the red giant branch is 
not well populated, metal-rich stars, i.e., ${\rm [Fe/H] \ga -0.5}$, 
that occupy the red part of the CMD where no calibration is available, 
might make the tip of the red giant branch less well identifiable.
Our simple color cut accomplishes this goal, but will lead to 
different overall luminosity functions for different metallicity
distributions. This is not a problem, provided that the TRGB 
feature itself is clearly visible.
The mean metallicities of halo stellar populations of all our 
sample galaxies are at or below ${\rm [Fe/H]\la\,-0.6}$ (see the 
second paper of this series for more details), so any contamination 
from metal-rich stars is expected to be small.

We used two different methods to detect and to measure the TRGB 
I-band magnitude from the luminosity function of a stellar population. 
In the following subsections we present briefly our methodology. 

\subsection{ Edge Detection}

The most widely used procedure to estimate the TRGB I-band magnitude 
uses the abrupt termination of the RGB luminosity function at the 
TRGB magnitude. The TRGB discontinuity causes a peak in the first 
derivative of the observed stellar magnitude distribution (Lee et al. 
1993; Madore \& Freedman 1995). Sakai et al. (1996) update this method 
by Gaussian-smoothing the luminosity function to avoid the arbitrary 
choice of the binning size, and use a continuous edge detection 
function.

Due to the power-law form of the RGB luminosity function, we use a 
kernel-smoothed logarithmic luminosity distribution function filtered 
by a Sobel kernel to determine the luminosity of the TRGB; the edge
detection function is:

\begin{equation} 
E(m)\,=\,\log_{10}(\Phi(m+\sigma_{m}))-
         \log_{10}(\Phi(m-\sigma_{m})).
\end{equation}

where $\sigma_{m}$ is the photometric error, estimated as the
mean of photometric errors shown in Fig.\,\ref{phot_error}, 
within a bin of $\pm 0.05$ magnitude about a magnitude $m$ 
(Sakai et al. 1996).
The edge detection filter essentially measures the slope of the 
luminosity function, and hence it is very sensitive to the noise; 
any sudden change in the signal is extremely amplified by the 
filtering scheme. To suppress any detection of statistically 
non-significant peaks in the first derivative of the luminosity 
function, we apply a weighting scheme to the edge detection 
filtering output (Sakai et al. 1997; Mendez et al. 2002). Each 
$E(m)$ is weighted by the Poisson noise (i.e.,$\Phi(m)^{1/2}$). 

Cioni et al. (2000) have found the edge detection technique to be
a biased estimator of the TRGB magnitude. However, the bias depends 
on the photometric errors and the amount of smoothing in the 
luminosity function at the location of the TRGB. For our data set, 
the bias is less than $0.03-0.04$ magnitude (Cioni et al. 2000),
much smaller than the random errors; therefore the distance modulus 
to the galaxy sample we derive using the edge detection technique 
would not be significantly affected in a comparison with the maximum
likelihood (see below).

To estimate the statistical uncertainty in measuring the $I$-band 
TRGB magnitude, we use bootstrap resampling of the data to generate 
a large number of samples drawn from the original data set to simulate 
the act of observing multiple times. Using this procedure to quantify 
the formal errors in measuring the TRGB has the advantage of using 
the empirical distribution function derived from the data itself, 
rather than a particular functional form (e.g., Gaussian) for the
photometric errors. 
 
For each galaxy in our sample, we perform the bootstrap resampling 
and calculate the statistic of interest $N\sim\,n\times\,[\log(n)]^2$ 
times, where $n$ is the number of observed stars within the magnitude 
range where the luminosity function model is valid in the case of
maximum likelihood analysis, and the number of stars used to construct 
the luminosity function in the case of Sobel edge detection analysis. 
To this extent the bootstrap version of a statistic's sampling 
distribution matches the asymptotic sampling distribution (see Babu 
\& Feigelson 1996 and Efron \& Tibshirani 1986 for reviews of the 
theory and application of this technique). In these bootstrap samples, 
the standard deviation of the I-band TRGB distribution is taken as 
the random error in the measurement of the TRGB magnitude. 
This procedure allows a more robust estimate of the statistical 
errors than using the full-width at half maximum of the observed peak 
in the first luminosity function derivative, as usually done (Sakai 
et al. 1996; see also Mendez et al. 2002 and Cioni et al. 2001).

\subsection{ Maximum Likelihood Analysis}

An alternative to filtering is to fit models to the data using the
technique of maximum likelihood; this avoids concerns over binning
and kernel smoothing. While theory
does not provide strong guidance on the exact functional form of the
luminosity function near the TRGB, a simple two-power-law model appears
to provide a good empirical fit over a restricted range of
magnitudes near the TRGB itself. The TRGB itself is identified
as the break between the two power laws. 

In this technique, the underlying luminosity function is treated as a
probability distribution function (PDF) and we compare
relative likelihoods of drawing the observed data set from the model 
PDFs as we vary the model parameters. An advantage of this procedure 
is that no data binning is needed. In practice we have binned the both 
the probability distributions and the data in intervals of 0.002 mag, 
as it simplifies the fitting procedure. 

Given a model of the stellar distribution on the RGB, 
$\Phi(m | {\bf \theta})$k, where $\theta$ is the set of free parameters 
characterizing the model, the likelihood that the observed data set is 
drawn from this model is given simply by the product of individual 
probabilities that star $i$ has an  observed magnitude $m_i$ within a 
range $dm$:

\begin{equation}
{\cal L(\theta)} = \prod_{i} P_{i}({\bf \theta}) 
           \propto \prod_{i} \Phi(m_i | {\bf \theta})
\end{equation}

where $P_i$ is the individual probability defined as the ratio of the 
number of stars actually {\it{observed}} at a given magnitude by the 
total number of stars expected given a luminosity function model 
$\Phi(m | {\bf \theta})$.

Photometric errors will affect the magnitude distribution of the observed 
stars. To account for this, we convolve the intrinsic luminosity function 
models with an appropriate {\it{broadening}} function, $\Sigma(m)$, to obtain 
the distribution of {\it{observed}} magnitudes. This function describes the 
probability that a star with an intrinsic magnitude $m_{int}$ is observed 
to have the magnitude $m_{obs}=m_{int}+m$. Thus the distribution of the 
observed magnitudes is:

\begin{equation} 
\Phi(m_{i}) =  \int g(m|\theta) \Sigma(m_{i},m)dm   
\label{dist}
\end{equation} 

where $g(m|\theta)$ is the intrinsic luminosity function model. 

We represent the measurement error by a Gaussian function with a 
total dispersion $\sigma_{m}$. That is, we have

\begin{equation} 
 \Sigma(m_{i},m)\,=\,\frac{1}{(2\pi)^{1/2}\,\sigma_{m_{i}}}
                  \exp\left[-\frac{(m-m_{i})^2}{2\sigma_{m_{i}}^2}\right]
\end{equation} 

where $\sigma_{m_{i}}$ is the photometric error at the magnitude 
$m_{i}$. The probability of observing a star of a magnitude $m_i$ 
in a magnitude range $dm$ is simply $g(m_i|\theta)\,dm$. 
The photometric errors smear the probability to the shape of the 
error function, centered at $m_i$; then the probability becomes
$\Sigma(m_{i},m)\,g(m_i|\theta)\,dm$. 
The photometric errors were determined from artificial star 
tests. We take the output of these simulations in 0.5 mag intervals 
for input stars of a fixed color V-I = 1.0. In each 0.5 mag interval, 
we determine the mean and standard deviation of the residuals of 
recovered minus input I-band magnitudes. For smearing the model 
luminosity functions, we use a set of Gaussian kernels with the 
mean and standard deviation that were determined from the 
simulations and with an overall normalization set by the fractional
completeness in that 0.5 mag interval. 
The model luminosity functions are convolved with these kernels 
prior to estimating the logarithm of the likelihood.

The maximum likelihood fitting is restricted to the magnitude range 
where the effects of both the photometric errors and the incompleteness 
are not large. To avoid contamination from metal-rich stars that may 
bias the estimate of the TRGB location, we consider only blue stars, 
i.e., stars within the color range ${\rm -2<(V-I)<2}$ (in practice
there are few stars bluer than $V-I=0.5$). 
Ideally, it would be best to fit a curved TRGB to a locus of
RGB models of all metallicities. However, in practice the 
resolution of existing isochrones near the TRGB is insufficient 
for this purpose, and also the isochrones themselves have been
constructed via a series of interpolations that become increasingly
uncertain near the TRGB. For this reason, the TRGB distance
indicator remains an empirical test: a search for a clear jump
or inflection in the luminosity function. 

To approximate the intrinsic luminosity function $g(m|\theta)$, we 
use a broken power-law model (Cioni et al. 2000; Mendez et al. 2002; 
Gregg et al. 2004). The free parameters characterizing the intrinsic 
model are the logarithm of the amplitude of the discontinuity at the 
TRGB, the apparent magnitude of the TRGB break, and the slopes of 
luminosity function on either side of the break.

We used both the Nelder-Mead (1965) simplex algorithm and 
``simulated annealing'' (e.g., Metropolis, 1953; Press et al. 1992) to 
determine the best-fit model. Results of the two methods were similar,
and well within the uncertainties derived from the subsequent error 
analysis.

To determine the uncertainties in the TRGB magnitude, we have 
stochastically sampled the full model parameter space by building a 
Monte-Carlo Markov Chain (MCMC) using the Metropolis-Hastings update 
mechanism. There is a growing body of literature on this technique
(e.g., Knox, Christensen \& Skordkis 2001; Verde et al. 2003),
which provides an
efficient way to characterize the likelihood manifolds in a
multi-dimensional parameter space. 
The Metropolis-Hastings MCMC algorithm is a restricted random walk
through parameter space. At each step in the random walk,
magnitude and direction of the next step is drawn from a normalized 
``proposal density'' $q(y,x)$, where the new state $y$ depends only
on the current state $x$. The candidate $y$ is accepted or rejected
with a probability 
\begin{equation}
{\rm min} \left(1,\frac{p(y)}{p(x)}\frac{q(y,x)}{q(x,y)}\right),
\end{equation}
where $p(y)/p(x)$ is the likelihood ratio for state $x$ relative to state $y$. 
After a period of
``burn-in,'' the distribution of locations in parameter space sampled
by the chain converges to the ``posterior distribution'' $p$. Thus likelihood
contours can be constructed simply by drawing contours of the density of 
points sampled by the chain. 

For our proposal distribution, we have used a Gaussian with $\sigma =
0.1$ in all of the fit parameters. We have also restricted the chains
to operate within a ``reasonable'' range of parameter space:  break
amplitudes within a factor of 100 of the best fit, TRGB magnitudes
within 1.5-2 mag of the best fit, and LF slopes between $-5$ and
$+0.2$.  The chains were burned in for 1000 steps, and then 10000 steps
were used to sample the posterior distributions. The probability
distributions thus derived are shown in the lower-right panels of
Figures 3-6.  We have verified that the results are insensitive to
small changes in the range of magnitudes being fit and the parameters
used to define the MCMC. The confidence intervals (and indeed
the best fit values) are nevertheless subject to a number of implicit 
prior assumptions. These include the assumption that the distribution
can be adequately described by two power laws with a break, 
over the full magnitude range being fit. In practice, we do see
small but significant differences in the confidence intervals 
as we change the magnitude limits and color limits. However shifts
of 0.2 mag in these limits in all of our tests shift the peak
of the likelihood distribution by less than the $1\sigma$ widths
show in Figs. 3-7.

\section{ TRGB Distances}
\label{dist}

As a first step to investigate the properties of field stellar populations 
in spiral galaxy halos, we have determined distances to a sample of galaxies.
A summary of TRGB distances and uncertainties is presented in 
Table \ref{gal_dist}. 
Color-magnitude diagrams, luminosity functions, and other properties for 
our galaxy sample are presented in the next subsections.
The best fitting luminosity function model, $\Phi(m | {\bf \hat{\theta}})$, 
where $\hat{\theta}$ is the free parameter set that maximize the likelihood, 
is overplotted on the observed luminosity function for each galaxy. Note 
that the smoothing by the photometric errors, accounted for in the maximum 
likelihood analysis, converts the abrupt break at the TRGB magnitude into 
a continuous RGB star distribution near the TRGB.    

Once the I-band magnitude of the TRGB is measured, we use the semi-empirical 
calibration given by Lee et al. (1993) to determine the distance modulus, 
starting from the relation:

\begin{equation}
 (m-M)_{I,o}\,=\,I_{TRGB}-M_{bol,TRGB}+BC_{I}
\end{equation}

The bolometric magnitude, M$_{bol,TRGB}$, and the bolometric correction,
BC$_{I}$, are function of the metallicity and the color of RGB stars
(Da Costa \& Armandroff 1990): 

\begin{equation}
M_{bol,TRGB}=-0.19\times[Fe/H]-3.81 
\end{equation}
\begin{equation}
BC_{I}=0.881-0.243\times(V-I)_{TRGB}
\end{equation}

where $(V-I)_{TRGB}$ is the color of stars at the TRGB magnitude.
The metallicity [Fe/H] is function of the RGB star colors (Lee et al. 1993): 

\begin{equation}
[Fe/H]=-12.65+12.6\times(V-I)_{-3.5}-3.3\times(V-I)^{2}_{-3.5} 
\end{equation}

where $(V-I)_{-3.5}$ is measured at an absolute I-band magnitude of 
$-3.5$ mag. To measure $(V-I)_{TRGB}$, we calculate the color distribution 
of stars within $I_{TRGB}\pm\,0.1$, and fit a Gaussian to the distribution. 
A similar procedure was repeated for the simulated bootstrap stars to 
measure the uncertainties, and a similar procedure was adopted to measure 
$(V-I)_{-3.5}$. The absolute I-band magnitudes are given for each galaxy in 
Table \ref{gal_dist}.

We assume that internal extinction is negligible as we are dealing with 
halo regions that are most likely dust free. To correct for the foreground 
extinction, we use the 100$\mu$m DIRBE/IRAS all-sky map of 
Schlegel et al. (1998).

\subsection{NGC 253}

NGC\,253 is a well studied nearly edge-on Sc-type L$_{*}$ galaxy in the 
Sculptor group; it is an archetypal starburst galaxy, and is the closest 
edge-on infrared luminous galaxy (Radovich et al. 2001). 
Extended soft X-ray emission from a hot gaseous component was detected in 
the halo of NGC\,253. This X-ray emission seems to be powered by a large-scale 
outflow, driven by the nuclear starburst activity (Pietsch et al. 2000; 
Strickland et al. 2002).

\begin{figure}
\includegraphics[height=3.5in]{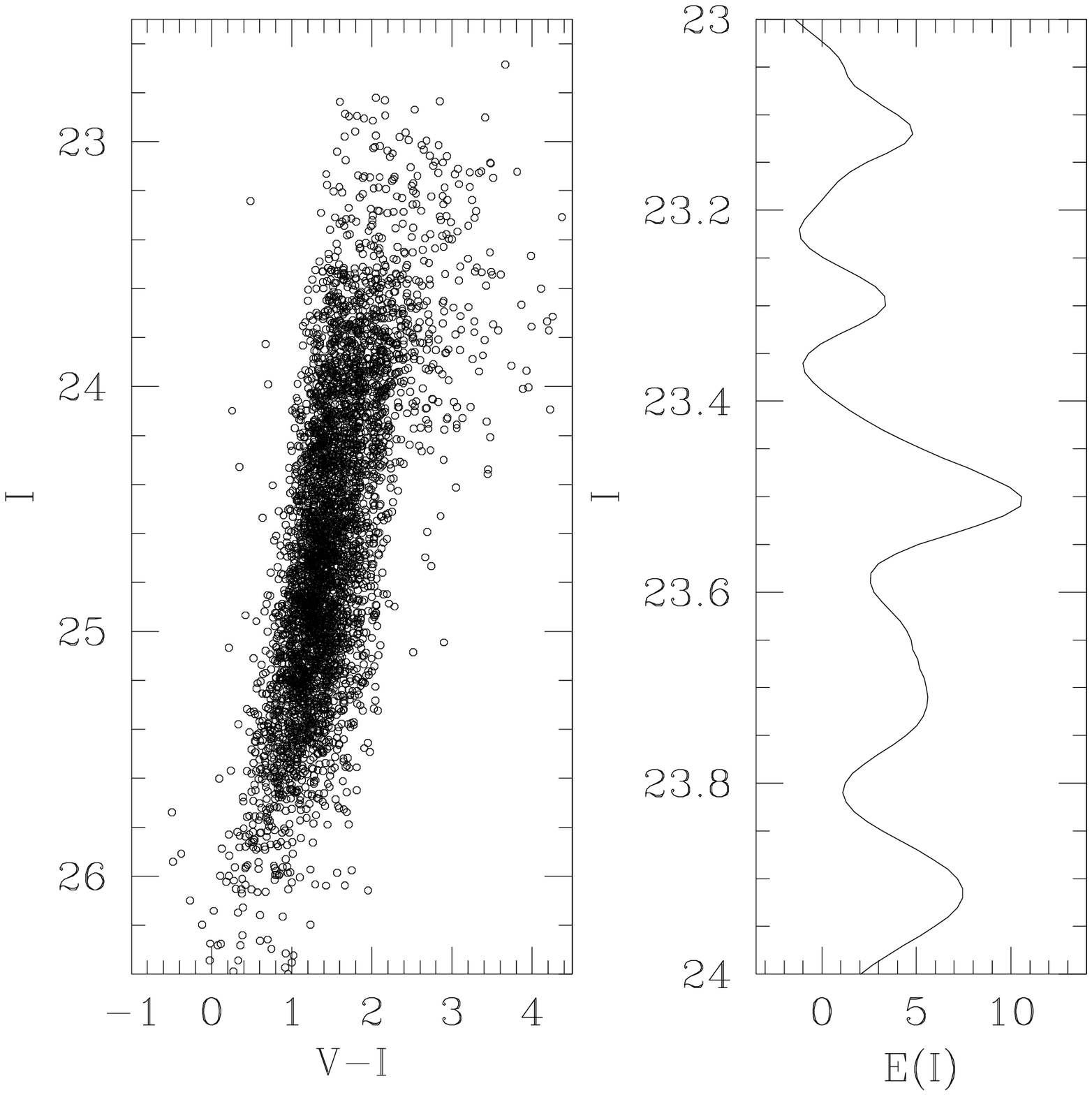}
\includegraphics[height=3.5in]{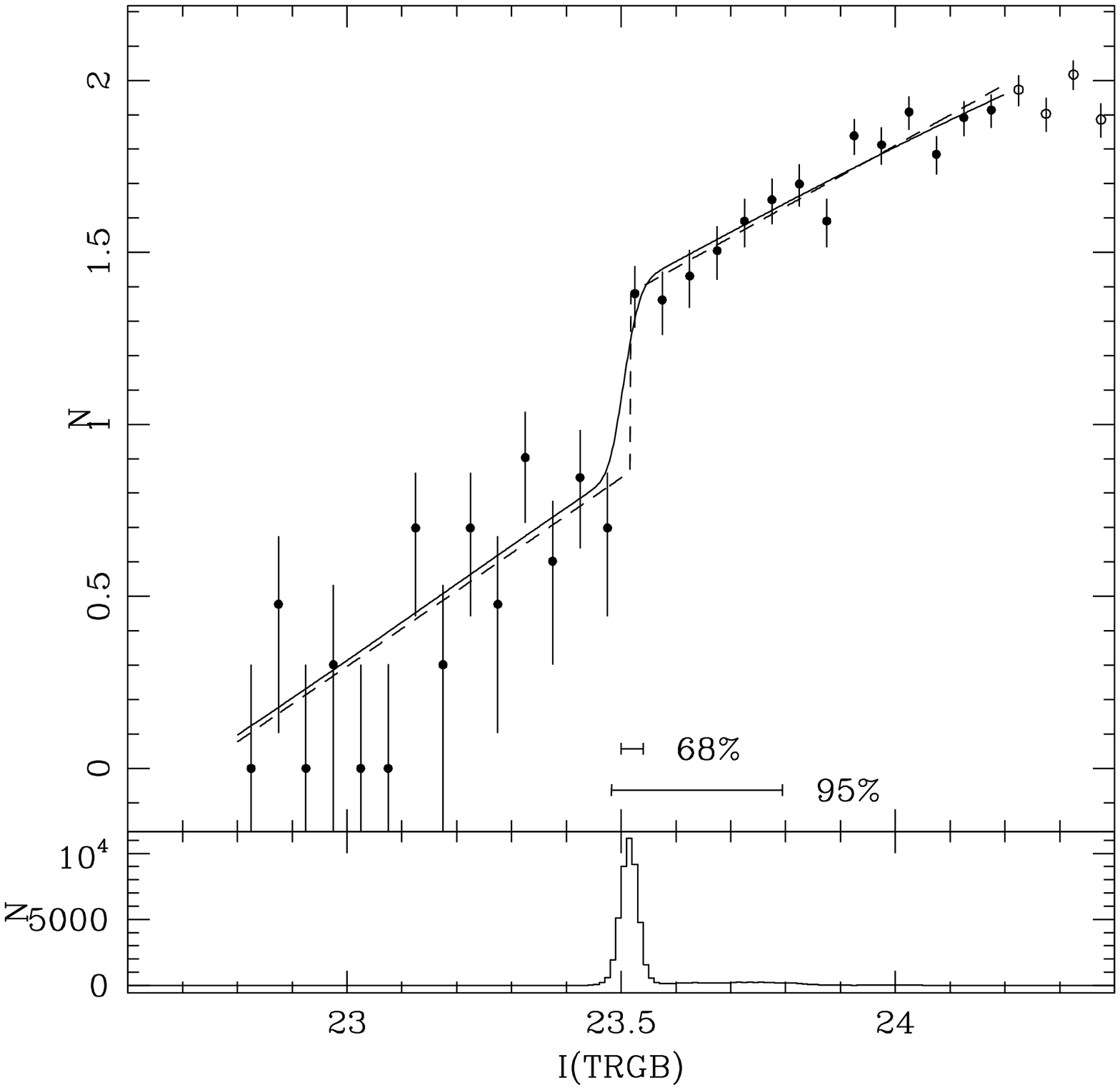}
\caption{Left: color-magnitude diagram for stars observed in the
NGC~253 halo field. The photometry has been corrected for foreground 
reddening. Center: the output of the weighted logarithmic 
edge detection filtering function.
Right: Luminosity function on a logarithmic scale. The data have
been modeled with two power laws using a maximum-likelihood fit.
Overplotted are the input best fit model (dashed line) and the 
model after convolution with the magnitude errors and application
of incompleteness (solid line).
The bottom panel shows the posterior probability distribution of 
the I-band TRGB magnitude derived from the Monte-Carlo Markov Chain 
analysis including the full range of the luminosity function free 
parameters.  The extension of the 95\% confidence toward fainter 
TRGB magnitudes interval represents a family of solutions that have 
steeper bright-end power-law slopes and smaller break-amplitudes
than the best-fit model shown here.}
\label{ngc253}
\end{figure}

The left panel of Fig.\,\ref{ngc253} displays the color-magnitude 
diagram of the observed halo field. The stars in the halo are predominantly 
red; the main feature of the color-magnitude diagram is the RGB structure, 
which finishes in a well-defined TRGB. The red giant branch is prominent 
and wide, indicating a large spread in the metallicity distribution of the halo 
stellar population. The stars above the TRGB and extending up to $V-I\sim\,4$ 
may be Asymptotic Giant Branch (AGB) stars covering a range of ages and 
metallicities. 

The color-magnitude diagram shows no indication of the presence of blue 
early-type main sequence stars in the observed halo field of NGC\,253, as 
was reported previously by Comer\'{o}n et al. (2001). 
The reason for this may be that the small number of blue stars detected 
toward the halo of NGC~253 are distributed over a large field. 
Wide field and deep imaging are needed to investigate in detail the star 
formation history in the halo of NGC\,253, and its potential link with 
the observed galactic outflow. 

In the panel next to the color--magnitude diagram, Fig. \ref{ngc253} shows 
the results of the edge 
detection filtering of the I-band luminosity function. The right panel 
of Fig.\,\ref{ngc253} shows the logarithmic luminosity function.  
The power-law distribution of the upper RGB stars is obvious, and has 
a clear break at $I_{TRGB}\sim\,23.5\,$mag. Overplotted are the 
maximum-likelihood best fit model before and after convolution with 
the photometric errors and application of incompleteness, shown as 
dashed and solid lines respectively. The bottom panel shows the 
posterior probability distribution of the TRGB magnitude derived from 
the Monte-Carlo Markov Chain analysis. Also shown are the $68\%$ and 
$95\%$ confidence intervals. 
Both the Sobel edge detector and the maximum likelihood model agree on 
the location of the I-band TRGB magnitude. For the maximum likelihood 
analysis, with a foreground reddening of E(B-V)=0.02\,mag and fitting 
data in the range $22.8<I<24.2$, we derive that the TRGB is located at 
$I_{TRGB}\,=\,23.5$ mag. The Monte-Carlo Markov Chain analysis gives 
that the $68\%$ confidence interval spans the range $23.50<I<23.54$ 
and the $95\%$ confidence interval is $23.48<I<23.79$. The faint 
extention of the $95\%$ confidence level is quite large, when the
TRGB is well defined, as there are a variety of models with different 
amplitudes for the jump and slopes above and below the TRGB that are 
consistent with the data to within the $95\%$ confidence interval.
There is a family of solutions with a steep powerlaw at the bright 
end and a much smaller jump than the best-fit shown in the plot.
The edge detection method locates the TRGB I-band magnitude at 
$I_{TRGB}\,=\,23.51$ mag with a bootstrap uncertainty of $\pm0.07$mag.
Following the procedure described in \S~\ref{dist}, we find a 
Population II distance modulus 
$(m-M)_{\circ} = 27.59 \pm 0.06$(random)$\pm 0.16$(systematic). 
This result is consistent with NGC~253 being a member of the nearby 
Sculptor group, located at the far side of the group; however, it is 
significantly larger than previously reported values derived using the 
globular cluster luminosity function (Blecha 1986), and the brightest 
halo (Davidge \& Pritchet 1990) and disk (Davidge, Le F\`evre, \& Clark 
1991) stars.

\subsection{ NGC 4244}

The edge-on galaxy NGC 4244 is a late-type (Scd) spiral; its total B-band 
magnitude is almost a magnitude below the peak of the field Sc luminosity 
function, making this galaxy less massive than a typical Sc.
The radial light distribution is well represented by an exponential profile
(Olling 1996), with almost no dark lanes (i.e., internal extinction is not 
important).

\begin{figure}
\includegraphics[height=3.5in]{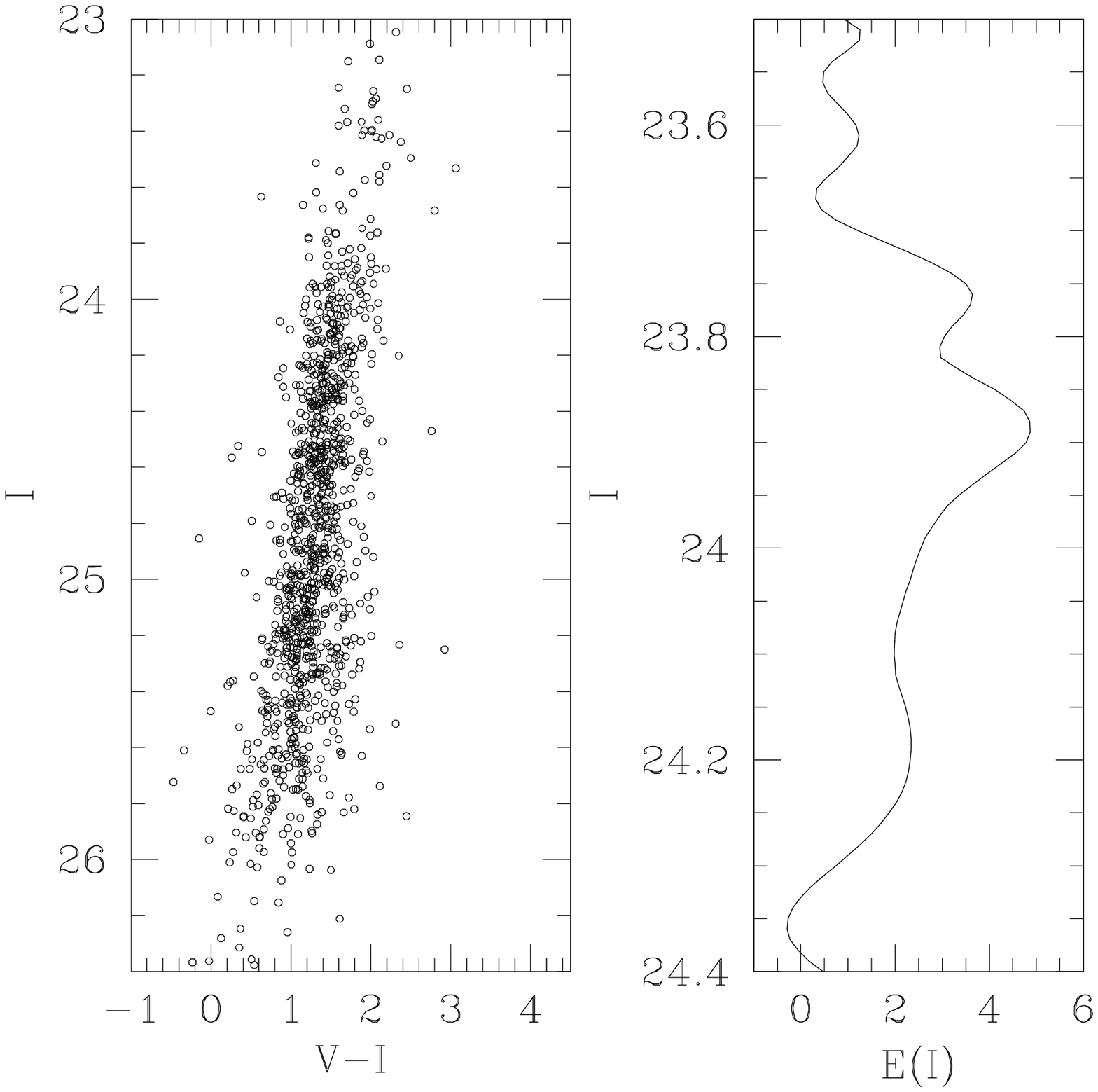}
\includegraphics[height=3.5in]{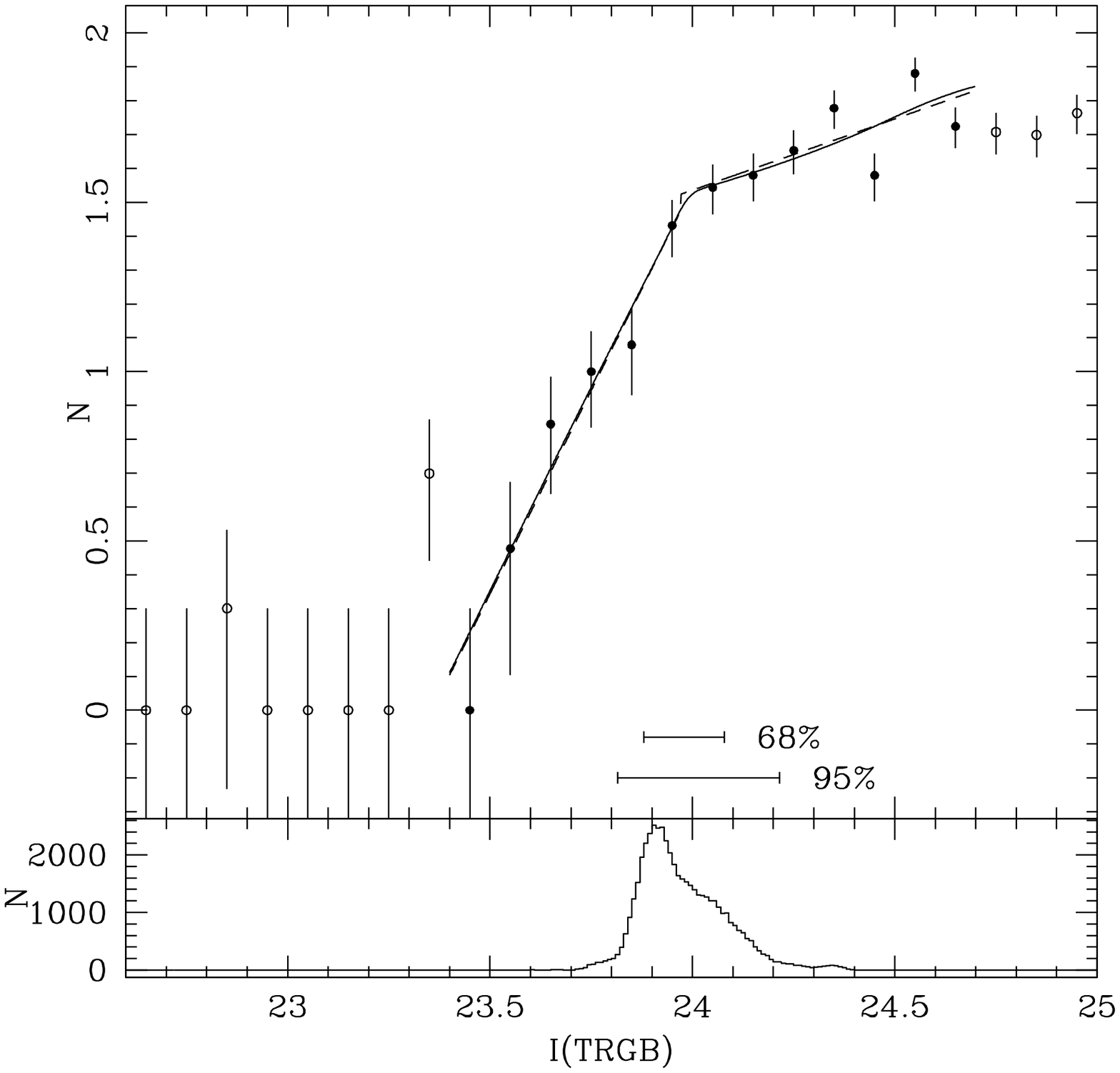}
\caption{Similar to Fig.\,\ref{ngc253} but for NGC~4244}
\label{ngc4244}
\end{figure}

The color--magnitude diagram for the observed halo field, shown in the left
panel of Fig.\,\ref{ngc4244}, reveals a simple stellar population; the red 
giant branch is prominent and clearly dominates the halo light, and there 
seems to be no blue early-type stellar population in the halo of this galaxy. 
The red giant branch is relatively tight, indicating a tight metallicity 
distribution in the halo stars, with a clear tip at $I\,\approx\,23.9$ mag. 
The stars above the tip are likely AGB stars. Also shown is the output of 
the edge detection filtering of the I-band luminosity function. 

The upper right panel of Fig.\,\ref{ngc4244} shows the logarithmic 
luminosity function for all stars detected in the field. For NGC~4244,
the TRGB appears as an inflection in slope, rather than a clear 
discontinuity. The power-law luminosity distribution of stars on the 
red giant branch is clear down to the magnitude where the effect of 
incompleteness starts to be important; the break of the luminosity 
function due to the TRGB, at $I\,\approx\,23.9$, is obvious. 
Again both the edge detection method and the maximum likelihood model 
agree on the location of the TRGB I-band magnitude. Using the maximum 
likelihood analysis, and fitting data in the range $23.4<I<24.7$, we 
find that the TRGB is located at $I_{TRGB}\,=\,23.9$ mag. 
The Monte-Carlo Markov Chain analysis gives a $68\%$ confidence interval 
of $23.88<I<24.08$ and a $95\%$ confidence interval of $23.82<I<24.22$. 
The edge detection method locates the TRGB at $I_{TRGB}\,=\,23.88$ mag. 
Resampling simulations give uncertainties of $\pm 0.1$ mag for the edge 
detection method. With a foreground extinction of $E(B-V)=0.021\,$mag, 
we find the distance modulus to NGC~4244 to be 
$(m-M)_{\circ} = 27.88 \pm 0.06$(random)$\pm 0.16$(systematic),
corresponding to a heliocentric metric distance of $D=3.82\,$Mpc.

Our distance modulus is in agreement with the previously published 
distance modulus for NGC~4244 of $\mu=27.78$ using the Tully-Fisher 
relation (Aaronson et al. 1986), but significantly different from the 
distance derived by Karachentsev \& Drozdovsky (1998), who used the 
mean B-band magnitude of blue supergiant candidates to derive a 
distance modulus of $\mu=28.28$. This method is affected by large 
uncertainties (Rozanski \& Rowan-Robinson 1994 and references therein).  
A preprint by Tikhonov \& Galazutdinova (2005) analyzing a larger 
data set on this galaxy, including the field analysed in this paper,
appeared after our paper was submitted. Their derived distance modulus 
using the tip of the red giant branch technique is 28.16, which is 
consistent within our 2$\sigma$ confidence interval but not our 
1$\sigma$ confidence interval. The magnitude of the tip of the 
red giant branch derived by Tikhonov \& Galazutdinova (2005) is 
fainter than what is estimated here (see Fig. 3 of paper III of this 
series, and Fig. 4 of Tikhonov \& Galazutdinova 2005). On the other 
hand, their estimated mean stellar metallicity is lower than our 
estimate (see paper II of this series). Tikhonov \& Galazutdinova 
(2005) have not well documented their data reduction and the procedure 
they used to measure the distance modulus to NGC~4244 to trace back 
accurately the source of the observed discrepancy.

\subsection{ NGC  4945}

NGC\,4945 is an edge-on Sb galaxy, hosting a powerful nuclear starburst 
region, with a ring morphology (Moorwood et al. 1996), powering a galactic 
superwind (Heckman, Armus, \& Miley 1990). Similar to NGC~5128, the 
prominent galaxy of the group to which NGC~4945 is believed to 
belong, its optical 
image is marked by dust extinction in the nuclear regions (Marconi et al. 
2000; Lipari et al. 1997). 

\begin{figure}
\includegraphics[height=3.5in]{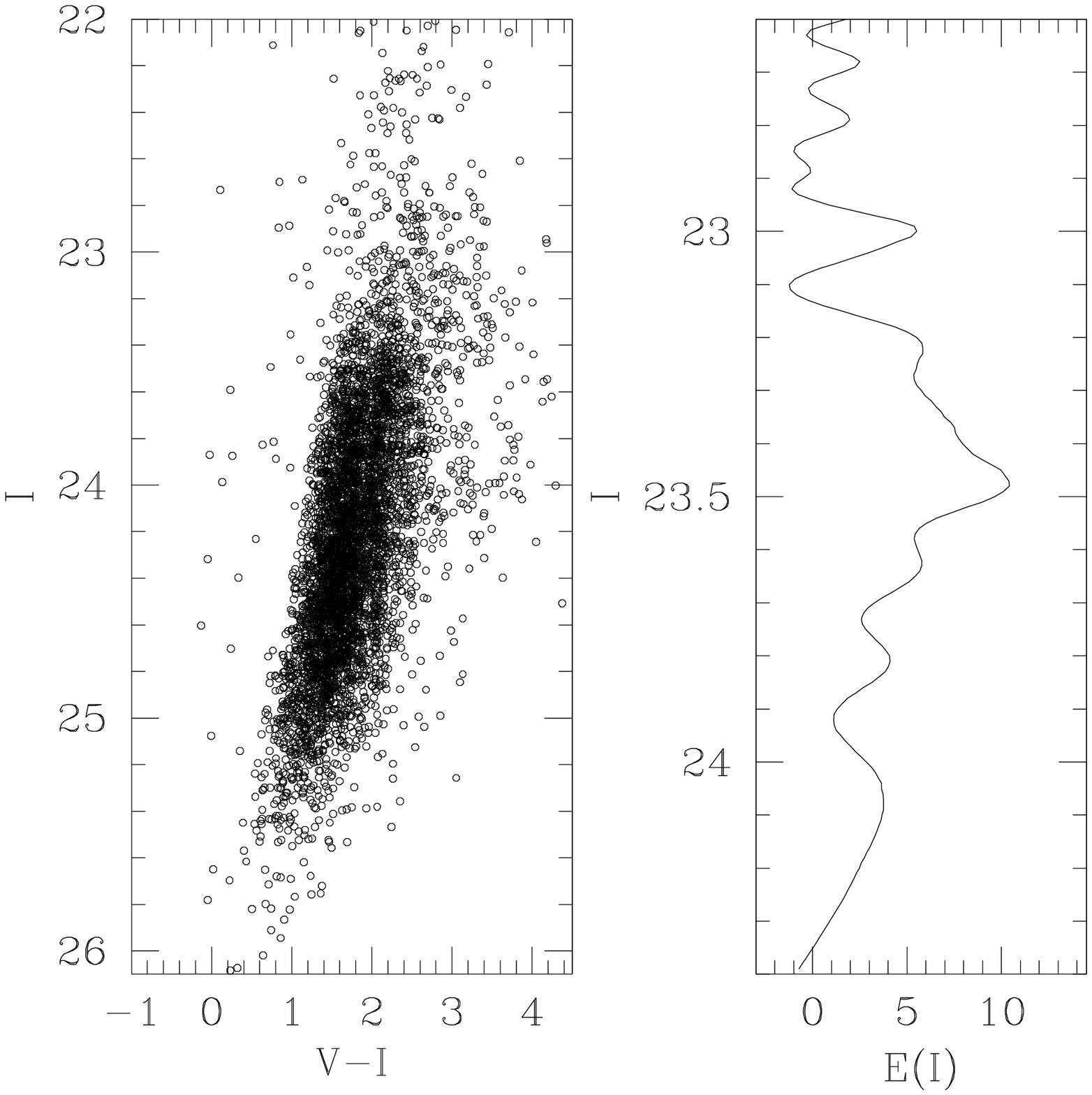}
\includegraphics[height=3.5in]{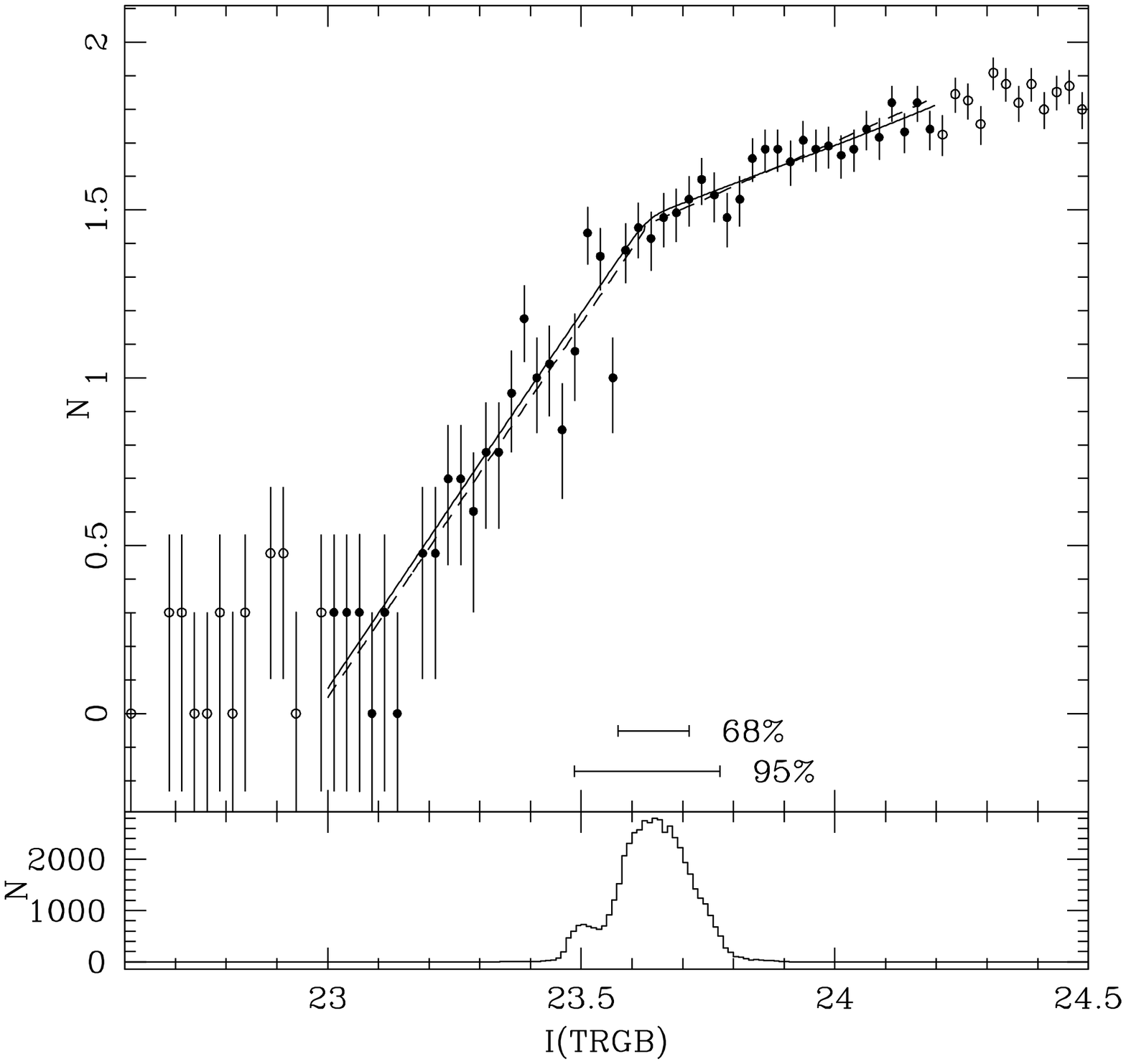}
\caption{Similar to Fig.\,\ref{ngc253} but for NGC~4945}
\label{ngc4945}
\end{figure}

The left panel of Fig.\,\ref{ngc4945} shows the color-magnitude diagram 
of the observed halo field stellar population. The red giant branch 
is wide and well populated, indicating a large spread of the halo stellar 
metallicities. Also shown is the output of the edge detection filtering 
of the I-band luminosity function. 

The right panel of Fig.\,\ref{ngc4945} shows the logarithmic I-band 
luminosity function, compared to the maximum-likelihood best model derived 
by fitting the data in the range $23<I<24.2$. Again the upper red giant 
branch power law distribution is evident with a break around 
$I\sim\,23.5-23.6$. 
The edge detection method and the maximum likelihood model disagree 
slightly on the location of the I-band TRGB. Using the maximum likelihood 
technique to fit the stellar luminosity function in the range $23<I<24.2$, 
we locate the TRGB magnitude at $I_{TRGB}\,=\,23.63$ mag, with a $68\%$ 
confidence interval of $23.57<I<23.71$ and a $95\%$ confidence interval 
of $23.49<I<23.77$. 
The edge detection method gives a brighter location of the TRGB at
$I_{TRGB}\,=\,23.48$ mag, with a bootstrap uncertainty of $\pm 0.06$mag.
The average location of the TRGB is $I_{TRGB}\,=\,23.55$ mag. 
Using this value with a foreground reddening of E(B-V)=0.18\,mag, we 
find the distance modulus for this galaxy to be
$(m-M)_{\circ} = 27.63 \pm 0.06$(random)$\pm 0.16$(systematic). 
This result is consistent with NGC~4945 being member of Centaurus A group,
which has a centroid at a distance modulus of $\mu=27.81\pm 0.11$ 
($3.66\pm0.19$Mpc; Karachentsev et al. 2002).

\subsection{ NGC 4258}

NGC~4258 is a nearby barred spiral galaxy hosting an obscured active nucleus 
and a massive nuclear black hole inside a highly inclined thin gaseous disk 
(Miyoshi et al. 1995; Maoz 1995).
Nuclear water maser sources were discovered moving in a Keplerian orbit around 
the central black hole of NGC 4258 (Watson \& Wallin 1994). By resolving these 
masers, measuring their acceleration (Greenhill et al. 1995), and monitoring 
their proper motions, Herrnstein et al. (1999) derived a distance modulus to 
the galaxy of $(m-M)_{\circ} = 29.28 \pm 0.09$.
NGC~4258 is one of the only two galaxies with a geometrical distance estimate, 
the other galaxy being the Large Magellanic Cloud, with a geometrical distance 
measurement based on the light echo of SN 1987A (Panagia et al. 1991). Thus 
NGC~4258 is a unique object for a cross-check of extragalactic distance 
techniques. 

Recently, Newman et al. (2001) used a sample of Cepheids in NGC~4258 to report
the distance modulus to this galaxy is $(m-M)_{\circ} = 29.47 \pm 0.09$ mag, 
with a distance modulus to the LMC of $(m-M)_{\circ} = 18.50$ mag. To do so, 
they used revised calibrations and methods for the HST Key Project on the 
Extragalactic Distance Scale. This distance is $1.2\,\sigma$ larger the the 
maser determination. To reconcile the geometrical and the Cepheid distances 
to NGC~4258, they argue that the true distance modulus to the LMC should be 
$18.31 \pm 0.11$(random)$\pm 0.17$(systematic), and should be used to calibrate 
the Cepheid brightness. 
Using the planetary nebulae luminosity function technique, Ciardullo et al. 
(2002) derived a distance modulus to NGC~4258 of 
$(m-M)_{\circ} = 29.42^{+0.2}_{-0.10}$. Similar to Newman et al (2001), they 
argue that the short distance scale to the LMC should be the correct one. 
On other hand, Caputo et al. (2002) argue that the theoretical metallicity 
correction, as suggested by pulsation models, is $\sim\,0.2$ mag lower 
than the empirical correction adopted by Newman et al. (2001). Then if 
this theoretical correction is applied to account for the chemical abundance 
differences between the LMC and NGC~4258, it may reconcile the Cepheid 
and the maser distances to NGC~4258, with the {\it{long distance scale}} 
to the LMC, a scale supported by different empirical non-Cepheid methods 
(Panagia et al. 1991; Sakai et al. 2000; Cioni et al. 2000; Groenewegen 
\& Salaris 2001). This debate raises again the well-known controversy between 
the short and long distance scales to the LMC.

We take advantage of our halo field observations to estimate the distance 
modulus to NGC~4258 using the TRGB technique. Provided we restrict
the ourselves to $V-I < 2$, this method is relatively 
insensitive to metallicity. The TRGB luminosity calibration does not
depend on the assumed distance to the LMC, but rather is computed from
theory and tested and calibrated via observations of globular clusters
in our galaxy. Because NGC~4258 is metal rich and the LMC is relatively
metal poor, the TRGB distance to 
NGC~4258 is an important cross-check of the metallicity effect on the 
Cepheid-based extragalactic distance scale.

\begin{figure}
\includegraphics[height=3.5in]{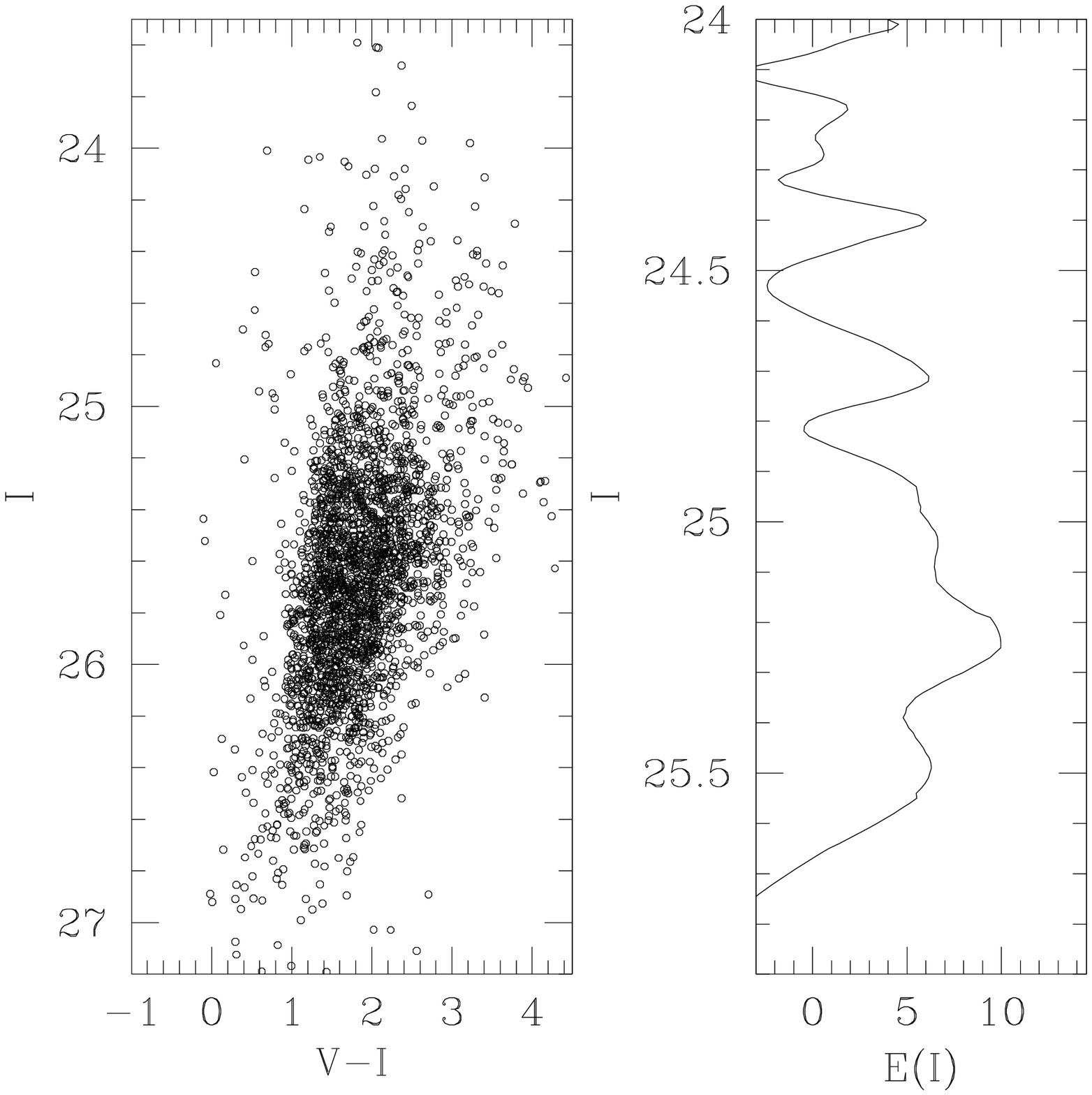}
\includegraphics[height=3.5in]{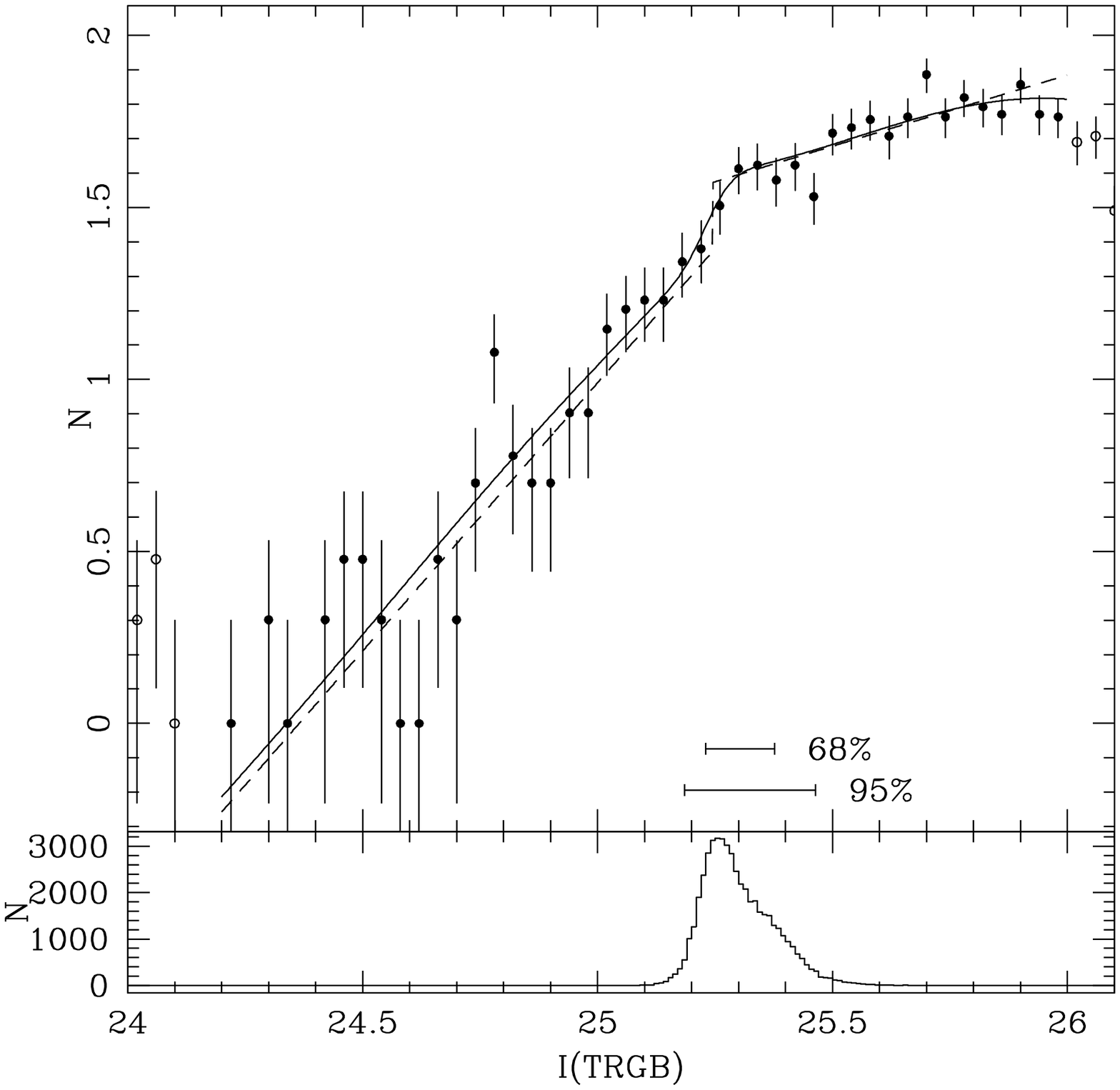}
\caption{Similar to Fig.\,\ref{ngc253} but for NGC~4258}
\label{ngc4258}
\end{figure}

The left panel of Fig.\,\ref{ngc4258} shows the color-magnitude diagram 
of the observed halo field stellar population. Again the halo stellar 
population is dominated by red giant branch stars. The red giant branch 
is wide, indicating a large spread of the stellar metallicity in the halo. 
Also plotted is the output of the edge detection filtering of the I-band 
luminosity function. 
The right panel of Fig.\,\ref{ngc4258} shows the logarithmic luminosity 
function for all stars observed in the field, compared to the maximum 
likelihood best fit model. The bottom panel shows the posterior probability 
distribution of the TRGB magnitude derived from the Monte-Carlo Markov 
Chain analysis.

Again both techniques agree on the location of the I-band TRGB magnitude.
For the maximum likelihood analysis, with a foreground foreground of 
E(B-V)=0.02\,mag and fitting data in the range $24.2<I<26$, we locate the 
TRGB at $I_{TRGB}\,=\,25.25$ mag. The Monte-Carlo Markov Chain analysis 
gives that the $68\%$ and $95\%$ confidence intervals are
$25.23<I<25.38$ and $25.19<I<25.46$, respectively.
The edge detection technique locates the TRGB at $I_{TRGB}\,=\,25.22$ mag 
with a bootstrap uncertainty of $\pm 0.09$ mag. Following the procedure 
described in \S~\ref{dist}, we find a Population II distance modulus for 
this galaxy to be 
$(m-M)_{\circ} = 29.32 \pm 0.09$(random)$\pm 0.15$(systematic), or a 
metric distance of D=$ 7.3 \pm 0.3$ Mpc. This is in superb agreement 
with both the maser and the metallicity-corrected Cepheid distances to 
the galaxy. Our distance modulus argue for the {\it{long distance scale}} 
to the LMC, where the true distance modulus to the LMC is close to 
$18.50$ mag.
The results provide support for Hubble constant 
$H_{\circ}=72 \pm 8$ km s$^{-1}$ Mpc$^{-1}$, inferred by Freedman et al. 
(2001) and based on the long distance scale to the LMC, in contrast to 
other recent suggestions (Maoz et al. 1999; Ciardullo et al. 2002).

\section{ Conclusion}
We report distance measurements using data obtained from a new program 
of deep HST imaging of the halos of nearby bright, edge-on spiral galaxies.

We have used the TRGB as a distance indicator for four nearby galaxies, 
NGC~253, NGC~4244, NGC~4945, and NGC~4258, observed with HST/WFPC2. 
We employ two methods, edge detection and maximum likelihood 
fitting, to measure the TRGB I-band magnitude from the luminosity 
function of RGB stars.  
Our results are in good agreement with prior distance measurements 
based on a variety of methods, including Cepheid variables.  However, 
we now have a common method of distance measurement for these galaxies.
The uncertainty in the estimate of the luminosity function break is 
calculated using the bootstrap resampling technique, which makes no 
assumption about the underlying distributions. 
Using our TRGB distance to NGC~4258, in combination with the galaxy's 
geometric and new metallicity-corrected distances, we argue that the 
long distance scale to the Large Magellanic Cloud is correct, and that 
the Freedman et al. (2001) Hubble constant should not be revised. 

Finally, increasing the number of galaxies beyond the Local Group with 
accurate TRGB measurements will certainly refine our understanding of 
the extragalactic distance scale, and will help making the systematic 
comparison between population I and population II distance indicators
 more meaningful.     

In future papers in this series, the same data set is used to derive 
abundance distributions in the halo stellar populations.

\acknowledgments

M.M would like to thank Bryan M\'{e}ndez for useful discussions. 
We acknowledge grants under HST-GO-9086 awarded by the Space Telescope 
Science Institute, which is operated by the Association of the 
Universities for Research in Astronomy, Inc., for NASA under contract 
NAS 5-26555


\clearpage

\begin{deluxetable}{lccccc}
\tablewidth{5.5in}
\tablecaption{Distance modulus for the galaxy sample. Columns: (1) Galaxy name;
(2) distance modulus estimates from the literature; (3) the method used to 
estimate the distance; (4) our distance modulus estimate; (5) Absolute I-band 
magnitude .}
\tablehead{
\colhead{Galaxy}&\colhead{$(m-M)_o^{old}$}&\colhead{method}
&\colhead{$(m-M)_o^{our}$}&\colhead{$M_{I}$}}
\tablecolumns{6}
\startdata

NGC~4258  & $29.28 \pm 0.06$   & Maser $^{1}$                     & $29.32 \pm 0.09$  & $-4.06 \pm 0.04$ & \\
          & $29.31 \pm 0.06$   & Cepheids + Z-correction$^{2}$    &                   &                  & \\
          & $29.47 \pm 0.12$   & Cepheids$^{3}$                   &                   &                  & \\
          & $29.42^{+0.07}_{-0.10}$  & PNLF$^{4}$                 &                   &                  & \\  
NGC~253   &   27.30      & Globulars LF$^{6,7}$                   & $27.59 \pm 0.06$  & $-4.09 \pm 0.02$ & \\
          & $\leq\,26.8$ & brightest halo and disk stars$^{8.9}$  &                   &                  & \\
NGC~4244  &   28.28      & brightest supergiants$^{10}$           & $27.88 \pm 0.06$  & $-4.02 \pm 0.03$ & \\
          &   27.78      & Tully-Fisher relation$^{11}$           &                   &                  & \\
NGC~4945  &   27.82      & Cen-A group centroid distance$^{12,5}$ & $27.63 \pm 0.06$  & $-4.06 \pm 0.02$ & \\

\enddata
\tablenotetext{1}{Herrnstein et al. 1999}
\tablenotetext{2}{Caputo et al. 2002}
\tablenotetext{3}{Newman et al. 2001}
\tablenotetext{4}{Ciardullo et al 2002}
\tablenotetext{5}{Ferrarese et al. 2000}
\tablenotetext{6}{Blecha 1986}
\tablenotetext{7}{Puche \& Carignan 1988}
\tablenotetext{8}{Davdige \& Pritchet 1990}
\tablenotetext{9}{Davidge et al. 1991}
\tablenotetext{10}{Karachentsev \& Drozdovsky 1998}
\tablenotetext{11}{Aaronson et al. 1986}
\tablenotetext{12}{Karachentsev et al. 2002}
\label{gal_dist}
\end{deluxetable}

\clearpage


\begin{thebibliography}{}

\bibitem[xxx]{xxx} Aaronson, M., Bothun, G., Mould, J., Huchra, J., Schommer, R.A., 
                   Cornell, M.E., 1986, \apj, 302, 536
\bibitem[xxx]{xxx} Aguilar, L., Hut, P., \& Ostriker, J.P., 1988, \apj, 335, 720
\bibitem[xxx]{xxx} Babu, G.J., \& Feigelson, E.D., ed. 1996, Astrostatistics 
                   interdisciplinary statistics
\bibitem[xxx]{xxx} Blecha, A., 1986, \aap, 154, 321
\bibitem[xxx]{xxx} Comer\'on, F., Torra, J., M\'endez, R.A., G\'omez, A.E., 
                   2001, \aap, 366, 796
\bibitem[xxx]{xxx} Caputo, F., Marconi, M., Musella, I., 2002, \apj, 566, 833        
\bibitem[xxx]{xxx} Carney, B.W., 1993, in The Globular cluster-Galaxy connection, 
                   eds. Smith, G.H., \& Brodie J.P., ASP Conf. Ser., 234
\bibitem[xxx]{xxx} Chiosi, C., Bertelli, G., \& Bressan, A., 1992, \araa, 
                   30, 235 
\bibitem[xxx]{xxx} Ciardullo, R., et al. , 2002, \apj, 577, 31             
\bibitem[xxx]{xxx} Cioni, M.-R.L, van den Marel, R.P., Loup, C., \& Habing, H.J.,
                   2000, \aap, 359, 601   
\bibitem[xxx]{xxx} Da Costa, G.S., \& Armandroff, T.E., 1990, \aj, 100, 162    
\bibitem[xxx]{xxx} Davidge, T.J., \& Pritchet, C.J., 1990, \aj, 100, 102
\bibitem[xxx]{xxx} Davidge, T.J., Le F\`evre, O., \& Clark, C.C., 1991, \apj, 
                   370, 559
\bibitem[xxx]{xxx} Dohm-Palmer, R.C., Helmi, A., Marrison, H., et al., 2001, 
                   \apj, 555, L37
\bibitem[xxx]{xxx} Dolphin, A.E., 2000, \pasp, 112, 1397 
\bibitem[xxx]{xxx} Durrell, P.R., Harris, W.E., \& Pritchet, C.J., 2004, \aj,
                   128, 260
\bibitem[xxx]{xxx} Efron, B., \& Tibshirani, R., 1986, Stat. Sci, 1, 54            
\bibitem[xxx]{xxx} Eggen, O., Lynden-Bell, D., \& Sandage A., 1962, \apj, 
                   136, 748
\bibitem[xxx]{xxx} Ferrarese, L., Mould, J.R., Kennicutt, R.C., Jr., et al., 
                   2000, \apjs, 128, 431 
\bibitem[xxx]{xxx} Freedman W.L., et al. 2001, \apj, 553, 47
\bibitem[xxx]{xxx} Freeman, K.C., 1987, \araa, 25, 603  
\bibitem[xxx]{xxx} Freeman, K.C., \& Bland-Hawthorn, J., 2002, \araa, 40, 487      
\bibitem[xxx]{xxx} Fruchter, A.S., \& Hook, R.N., 2002, \pasp, 114, 144
\bibitem[xxx]{xxx} Greenhill, L.J., Jiang, D.R., Moran, J.M., Reid, M.J., Lo, K.Y., 
                   Claussen, M.J., 1995, \apj, 440, 619
\bibitem[xxx]{xxx} Gregg, M.D., Ferguson, H.C., Minniti, D., Tanvir, A., 
                   Catchpole, R., 2004, \apj, 127, 1441
\bibitem[xxx]{xxx} Grillmair, C.J., Freeman, K.C., Irwin, M., Quinn, P.J., 1995, 
                   \aj, 109, 2553             
\bibitem[xxx]{xxx} Groenewegen, M.A.T., Salaris, M., 2001, \aap., 366, 752 
\bibitem[xxx]{xxx} Heckman T.M., Armus L., \& Miley G.K., 1990, \apjs, 74, 833
\bibitem[xxx]{xxx} Herrnstein J.R., et al., 1999, Nature, 400, 539
\bibitem[xxx]{xxx} Ho, L.C., Filippenko A.V., Sargent W.W, 1996, \apj, 462, 183
\bibitem[xxx]{xxx} Holtzman J.A. et al., 1995, PASP, 107, 1065 
\bibitem[xxx]{xxx} Ibata, R.A., Gilmore, G., \& Irwin M.J., 1994, Nature, 370, 194
\bibitem[xxx]{xxx} Ibata, R.A., Irwin M.J., Lewis, G., Ferguson, A.M.N., 
                   \& Tanvir, N., 2001, Nature, 412, 49
\bibitem[xxx]{xxx} Ivezic, Z., Goldston, J., Finlator, K., et al., 2000, \apj, 
                   120, 963                
\bibitem[xxx]{xxx} Karachentsev, I.D., \& Drozdovsky, I.O., 1998, \aaps, 131, 1    
\bibitem[xxx]{xxx} Karachentsev, I.D., Sharina, M.E., Dolphin, A.E., et al., 2002, 
                   \aap, 385, 21
\bibitem[xxx]{xxx} Knox, L., Christensen, N. \& Skordis, C., 2001, ApJ, 563, L95
\bibitem[xxx]{xxx} Krist, J., 2004, TinyTim manual, 
                   http://www.stsci.edu/software/tinytim/tinytim.pdf   
\bibitem[xxx]{xxx} Lee, M.G., Freedman, W.L., \& Madore, B.F. 1993, 
                   \apj, 417, 553
\bibitem[xxx]{xxx} Lipari, S., Tsvetanov, Z., Macchetto, F., 1997, \apjs, 111, 369  
\bibitem[xxx]{xxx} Madore, B.F.,  \& Freedman, W.L. 1995, \aj, 109, 1645  
\bibitem[xxx]{xxx} Marconi, A., Oliva, E., van der Werf, P.P., Maiolino, R., 
                   Schreier, E.J., Macchetto, F., Moorwood, A.F.M., 2000, \aap, 
                   357, 24                 
\bibitem[xxx]{xxx} Mendez, B., Davis, M., Moustakas, J., Newman, J., 
                   Madore, B.F., \& Freedman W.L., 2002, \apj, 124, 213
\bibitem[xxx]{xxx} Miyoshi, M., Moran, J., Herrnstein, J., Greenhill, L., 
                   Nakai, N., Diamond, P., \& Inoue, M., 1995, Nature, 373, 127
\bibitem[xxx]{xxx} Maoz, E., 1995, \apj, 455, 131            
\bibitem[xxx]{xxx} Maoz D., et al., 1999, Nature, 401, 351                     
\bibitem[xxx]{xxx} Metropolis, N., Rosenbluth, A. W., Rosenbluth, M. N., 
                   Tellar, A. H., \& Tellar, E., 1953, Journal of Chemical 
                   Physics, 21, 6.
\bibitem[xxx]{xxx} Moorwood, A.F.M., et al. 1996, \aap, 308, L1             
\bibitem[xxx]{xxx} Morrison, H.L., 1993, \aj, 106, 578     
\bibitem[xxx]{xxx} Nelder, J. A. \& Mead, R., 1965, Computer Journal, 7, 308
\bibitem[xxx]{xxx} Newman J.A., et al., 2001, \apj, 553, 562 
\bibitem[xxx]{xxx} Olling, R.P., 1996, \aj, 112, 481    
\bibitem[xxx]{xxx} Panagia, N., Gilmozzi, R., Macchetto, F., Adorf, H.-M., 
                   \& Kirshner, R.P., 1991, \apj, 380, L23       
\bibitem[xxx]{xxx} Perrett, K.M., Bridges, T.J., Hanes, D.A., et al., 2002, 
                   \aj, 123, 2490
\bibitem[xxx]{xxx} Pietsch, W., Vogler, A., Klein, U., \& Zinnecker, H.,
                   2000, \aap, 360, 24  
\bibitem[xxx]{xxx} Press, W.H., Teukolsky, S.A., Vetterling, W.T. 
                   \& Flannery, B.P., 1992, Numerical Recipes in C, Cambridge 
                   University Press, Cambridge.
\bibitem[xxx]{xxx} Puche, A., \& Carignan, C., 1988, \aj, 95, 1025
\bibitem[xxx]{xxx} Radovich, M., Kahanpaa, J., Lemke, D., 2001, \aap, 377, 73
\bibitem[xxx]{xxx} Renzin, A., 1992, in IAU Symp. 149, The Stellar Populations
                   in Galaxies, Ed. B. Barbuy \& A. Renzini (Dordrecht: Kluwer) 
                   235
\bibitem[xxx]{xxx} Rozanski R., \& Rowan-Robinson, M., 1994, \mnras, 271, 530
\bibitem[xxx]{xxx} Sakai, S., Madore, B.F., \& Freedman, W.L. 1996, 
                   \apj, 461, 713                  
\bibitem[xxx]{xxx} Sakai, S., Madore, B.F., Freedmen, W.L., Lauer, T.R., 
                   Ajhar, E.A., \& Baum, W.A., 1997, \apj, 478, 49 
\bibitem[xxx]{xxx} Sakai, S., Zaritsky, D., Kennicutt, R.C., Jr., 2000, \aj, 
                   119, 1197    
\bibitem[xxx]{xxx} Salaris, M., \& Cassisi, S.,  1997, \mnras, 289, 406 
\bibitem[xxx]{xxx} Salaris, M., \& Girardi, L., 2005, \mnras, 357, 669
\bibitem[xxx]{xxx} Searle, L., \& Zinn R., 1978, \apj, 225, 357
\bibitem[xxx]{xxx} Schlegel, D.J., Finkbeiner, D.P., \& Davis, M., 1998, 
                   \apj, 500, 525
\bibitem[xxx]{xxx} Strickland, D.K., Heckman, T.M., Weaver, K.A., Hoopes, C.G.,
                   Dahlem, M., 2002, \apj, 568, 689        
\bibitem[xxx]{xxx} Tikhonov, N.A., Galazutdinova, O.A., \& Drozdovsky, I.O.,
                   2005, \aap, 431, 127.
\bibitem[xxx]{xxx} Tikhonov, N.A., \& Galazutdinova, O.A., 2005, Astrofizica, 
                   in press, astro-ph/0503235.
\bibitem[xxx]{xxx} Watson, W.D., \& Wallin, B.K., 1994, \apj, 437, L35         
\bibitem[xxx]{xxx} Verde, L., et al., 2003, ApJS, 148, 195.
\bibitem[xxx]{xxx} Yanny, B., Newberg, H.J., Kent, S., et al., 2000, \apj, 
                   540, 825
\bibitem[xxx]{xxx} Zoccali, M., \& Piotto, G., 2000, \aap, 358, 943

\end{thebibliography}
\end{document}